\documentclass[twoside]{article}
\usepackage{fleqn,espcrc2}

\usepackage{graphicx}

\newcommand{\be}{\begin{equation}}
\newcommand{\ee}{\end{equation}}
\def\aprle{\buildrel < \over {_{\sim}}}

\newcommand{\AmS}{{\protect\the\textfont2
  A\kern-.1667em\lower.5ex\hbox{M}\kern-.125emS}}

\begin{document}
\title{\normalsize \hfill FISIST/20-2000/CFIF \\ [1 cm] \Large Seesaw neutrino masses and
mixing with extended democracy \thanks{Talk given at the {\em
Europhysics Neutrino Oscillation Workshop - Now 2000}, Conca
Spechiulla (Otranto)-Italy.}}

\author{F. R. Joaquim\address{Centro de F{\'{\i}}sica das Interac\c c\~oes
Fundamentais (CFIF)
\\
Instituto Superior T\'ecnico - Departamento de F{\'{\i}}sica \\Av. Rovisco Pais \\ P-1049-001 Lisboa, Portugal }%
 \thanks{Work supported by {\em Funda\c c\~ao para a Ci\^encia e a Tecnologia} under the grant PRAXIS XXI/BD/18219/98.} }

\begin{abstract}
In the context of a minimal extension of the Standard Model with
three extra heavy right-handed neutrinos, we propose a model for
neutrino masses and mixing based on the hipothesis of a complete
alignment of the lepton mass matrices in flavour space.
Considering a uniform quasi-democratic structure for these
matrices, we show that, in the presence of a highly hierarchical
right-handed neutrino mass spectrum, the effective neutrino mass
matrix, obtained through the seesaw mechanism, can reproduce all
the solutions of the solar neutrino problem.\vspace{1pc}
\end{abstract}

\maketitle All the enthusiasm around the Super-Kamiokande evidence
for atmospheric neutrino oscillations \cite{SK} has lead to a
great activity in the search for models \cite{review} which can
reproduce the solar and atmospheric neutrino data \cite{data}.
\\ \indent In Ref.~\cite{our} we have proposed a simple model for lepton masses
and mixing which fulfils these requirements.\\
 \indent As a starting point, we consider the existence of three heavy
right-handed neutrinos which, together with the seesaw mechanism,
will generate small neutrino masses. Furthermore, we assume that
the charged lepton, Dirac and right-handed neutrino mass matrices
($M_l$, $M_D$ and $M_R$ respectively) are aligned in flavour space
in a such a way that they are all ``democratic'', i.e., they are
proportional to a matrix whose elements are, in leading
approximation, all equal to one. We will refer to this assumption
as ``extended democracy''. In the context of the Standard Model
(SM) with three additional heavy right-handed neutrinos, the
lepton mass Lagrangian has the form:
\begin{eqnarray} -{\cal L}_{mass}=\bar{l}_{iL}\,(M_l)_{ij}\,l_{jR}
+\bar{\nu}_{iL}\,(M_D)_{ij}\, \nu_{jR}+ \nonumber \\
\quad+\frac{1}{2}\,\nu_{iR}^T \,C\,(M_R)_{ij}\,
\nu_{jR}+h.c.\,\,.\hspace{0.9cm} \label{L}
\end{eqnarray}

Following the hints of some Grand Unified Theories (GUTs), we
consider the mass spectrum of $M_D$ similar to the one of the
up-type quarks. \\ \indent Since the matrices $M_l$, $M_D$ and
$M_R$ are assumed to be, at leading order, proportional to the
democratic matrix $\Delta$, we write: \be M_k =c_k\,[\,
\Delta\,+\,P_k]\;,\quad k = l,D,R\;, \label{mat1} \ee \be
{\Delta}_{ij}=1\,\,\,,\,\,\,P_k={\rm diag}(0,a_k,b_k)\;,
\label{mat2} \ee
with $|\,a_k|\,,|\,b_k|\ll1$, so that all the
matrices are close to the democratic limit. From Eq.~(\ref{mat2})
one can see that the breaking of the extended democracy is small
and it has the same pattern for all the mass matrices.
 The effective neutrino mass matrix, $M_{\rm eff}=-{M_D}^T{M_R}^{-1}M_D$, is given by:
\be M_{\rm eff}=-c_{\rm eff}\,[\,\Delta + P_{\rm
eff}\,]\,,\,P_{\rm eff}={\rm diag}(0,x,y), \label{mef3} \ee
where
$x={a_D}^2/a_R$, $y={b_D}^2/b_R$ and $c_{\rm eff}={c_D}^2/c_R$. It
is interesting to notice that this matrix has the same general
form as the matrices $M_k$, i.e. the seesaw mechanism preserves
our $\it Ansatz$. This is a remarkable feature of the scheme we
propose in Eqs. (\ref{mat1}) and (\ref{mat2}).

Our next task is to constrain the parameters in the mass matrix
$M_{\rm eff}$ so as to satisfy the experimental bounds on the
values of the $\Delta m^{2}$'s and mixing angles.

The hierarchical structure of the eigenvalues of $M_l$ implies
$|a_l|, |b_l|\ll 1$. Since the matrix $\Delta$ can be diagonalized
as $F^T \Delta F={\rm diag}(0,\,0,\,3)$ with
\be F=\left(\begin{array}{ccc} ~~\frac{1}{\sqrt{2}} &
~~\frac{1}{\sqrt{6}} & ~~\frac{1}{\sqrt{3}} \\ -\frac{1}{\sqrt{2}}
& ~~\frac{1}{\sqrt{6}} & ~~\frac{1}{\sqrt{3}} \\ ~0  &
-\frac{2}{\sqrt{6}} & ~~\frac{1}{\sqrt{3}}
\end{array} \right)\,,
\label{F} \ee
the matrix $U_l$ that diagonalizes $M_l$ can be
written as $ U_l=FW$, where, due to the hierarchy $|a_l|\ll
|b_l|\ll 1$, the matrix $W$ is close to the unit matrix.
 Let us consider that all the parameters $a_k$ and $b_k$ in
Eqs.~(\ref{mat1}) and (\ref{mat2}) are real \footnote{ This
treatment can be extended to the most general case of complex mass
matrices, such as the special ones based on the hipothesis of
universal strength of Yukawa couplings (USY) \cite{our}.}. It is
instructive to analyse first the limit when the matrix $W$
coincides with the unit matrix, which corresponds to neglecting
$m_e$ and $m_{\mu}$. The parameters $a_k$, $b_k$ and $c_k$ are
related to the masses of charged leptons, up-type quarks and heavy
Majorana neutrinos through:
\begin{eqnarray}
a_l\simeq 6\frac{m_e}{m_\tau}\quad,\quad a_D\simeq
\frac{m_u}{m_t}\quad,\quad a_R\simeq 6\frac{M_1}{M_3}\,,\nonumber
\end{eqnarray}
\vspace{-0.5cm}
\begin{eqnarray}
b_l\simeq \frac{9}{2}\frac{m_\mu}{m_\tau}\;\;\;,\;\;\, b_D\simeq
\frac{9}{2}\frac{m_c}{m_t}\;\,\;,\,\;\;\; b_R\simeq
\frac{9}{2}\frac{M_2}{M_3}\,,\nonumber \end{eqnarray}
\vspace{-0.5cm}
\begin{eqnarray} |c_l|\simeq
\frac{m_\tau}{3}\;\,\;\;\,,\;\,\; |c_D|\simeq
\frac{m_t}{3}\;\,\;\,,\;\,\;\; |c_R|\simeq \frac{M_3}{3}\;,
\label{param1}
\end{eqnarray}
where $M_i$ ($i$=1,2,3) are the heavy right-handed neutrino
masses, which are the only free parameters in our model. The
effective mass matrix $\tilde{M}_{\rm eff}$, in the basis where
the charged leptons have been diagonalized, is then obtained from
Eq.~(\ref{mef3}) through the rotation by the matrix $F$:
$\tilde{M}_{\rm eff}=F^T M_{\rm eff} F\equiv-c_{\rm eff}
y\,\tilde{M}_0$, with $\tilde{M}_0$ given by:
\be\left(
\begin{array}{ccc}\;\;\;\frac{\varepsilon}{2} &
-\frac{\varepsilon}{2\sqrt{3}} & -\frac{\varepsilon}{\sqrt{6}}  \\
\vspace*{0.15cm} \,-\frac{\varepsilon}{2\sqrt{3}}&
\,\frac{2}{3}+\frac{\varepsilon}{6} &
\,-\frac{2}{3\sqrt{2}}+\frac{\varepsilon}{3\sqrt{2}}
\\ -\frac{\varepsilon}{\sqrt{6}} &
-\frac{2}{3\sqrt{2}}+\frac{\varepsilon}{3\sqrt{2}} &
\,\frac{1}{3}+\frac{\varepsilon}{3}+\delta \end{array} \right),
 \label{mef4} \ee
 where $\varepsilon=x/y$ and $\delta=3/y$. The condition
$\Delta m_{12}^2\equiv \Delta m_\odot^2 \ll \Delta m_{32}^2\equiv
\Delta m_{atm}^2$ requires $|\varepsilon|,|\delta|\ll 1$. Then,
the largest eigenvalue of the matrix $\tilde{M}_0$ is always close
to one. This leads to: \be M_2\simeq
\frac{3}{2}\frac{m_c^2}{\sqrt{\Delta m_{atm}^2}}\simeq 4\times
10^{10}~{\rm GeV}\,. \label{M2} \ee Moreover, we find:
\be M_1\simeq \frac{2}{\varepsilon} \frac{m_u^2}{\sqrt{\Delta
m_{atm}^2}}\quad,\quad M_3\simeq \frac{1}{\delta}
\frac{m_t^2}{\sqrt{\Delta m_{atm}^2}}\,. \label{M1M3} \ee It is
interesting to notice that the values of $M_2$ are always nearly
the same, which stems from the fact that they are related to
$\Delta m_{atm}^2$ ( \emph{cf.} Eq.~(\ref{M2})) and practically
independent from $\Delta m_\odot^2$ and $\theta_{12}$.
\\ \indent In the limit $|\varepsilon|\ll |\delta| \ll 1$
(relevant for the SMA solution of the solar neutrino problem), the
light neutrino masses are \begin{eqnarray} \{m_1,\, m_2,\,
m_3\}=-c_{\rm eff}\,y \left\{
\frac{\varepsilon}{2}\,,\frac{2}{3}\,\delta+\frac{\varepsilon}{2}\,,
1+\frac{\delta}{3}\right \}. \nonumber \end{eqnarray}
 The
diagonalization of $\tilde{M}_{\rm eff}$ yields \be \varepsilon
\simeq \sin 2\theta_{12}\sqrt{\frac{\Delta m_{21}^2}{\Delta
m_{32}^2}}\;\;,\;\;\delta \simeq \frac{3}{2}\sqrt{\frac{\Delta
m_{21}^2}{\Delta m_{32}^2}}\,, \label{param4} \ee \be \sin^2
2\theta_{23}\simeq \frac{8}{9}\left(1+\frac{2}{3}\delta\right)
\;\;,\;\; \sin\theta_{13}=-\frac{\varepsilon\delta}{3\sqrt{2}}\,.
\label{3} \ee
\begin{table*}[htb]
\caption{Results from exact numerical diagonalizations of
${\tilde{M}}_{\rm eff}$ corresponding to the four solutions of the
solar neutrino problem. The corrections due to non-zero $m_e$ and
$m_{\mu}$ were included. } \label{table:1}
\newcommand{\m}{\hphantom{$-$}}
\newcommand{\cc}[1]{\multicolumn{1}{c}{#1}}
\renewcommand{\tabcolsep}{2.0pc} 
\renewcommand{\arraystretch}{1.3} 

\begin{tabular}{@{}lcccc} \hline
& LMA &SMA&LOW&VO\\
\hline $a_R$ &$1.5\times10^{-9}$&
$8.5\times10^{-8}$&$1.9\times10^{-10}$ &$4.1\times10^{-9}$
\\$b_R$&$1.3\times10^{-5}$
&$2.2\times10^{-5}$ &$8.0\times10^{-8}$&
$3.9\times10^{-8}$\\
$M_3\,({\rm GeV})$&$1.3\times10^{16}$&$7.6\times10^{15}$
&$2.0\times10^{18}$&$4.3\times10^{18}$\\
\hline $M_1\,({\rm GeV})$&$3.2\times10^6$ &$1.1\times10^8$&
$6.3\times10^7$&$2.9\times10^9$ \\
$M_2\,({\rm GeV})$&$3.8\times10^{10}$&$3.7\times10^{10}$
&$3.6\times10^{10}$&$3.8\times10^{10}$ \\
$\Delta m_{12}^2\,{\rm (eV)^2}$ &$5.36\times10^{-5}$
&$7.25\times10^{-6}$&$1.15\times10^{-7}$ &$1.02\times10^{-10}$
\\$\Delta m_{23}^2\,{\rm (eV)^2}$&$3.94\times10^{-3}$ &$4.15\times10^{-3}$
&$4.35\times10^{-3}$ &$3.96\times10^{-3}$
\\$\sin^2 2\theta_{12}$&0.95 &$5.1\times10^{-3}$&0.999&0.70
\\$\sin^2 2\theta_{23}$&0.95&0.96&0.94&0.94\\
$U_{e3}$&$3.27\times10^{-3}$&$2.27\times10^{-3}$
&$2.29\times10^{-3}$&$2.29\times10^{-3}$\\ \hline
\end{tabular}
\end{table*}
Similarly, we can derive the following expressions for the
eigenvalues of $\tilde{M}_{\rm eff}$ in the case $|\delta|\aprle
|\varepsilon| \ll 1$ (relevant for the VO, LOW and LMA solutions
of the solar neutrino problem): \begin{eqnarray}  \{m_1,\, m_2,\,
m_3\}= \nonumber\end{eqnarray}
\begin{eqnarray}\quad\quad\quad\,
-c_{\rm eff} y\left\{
\frac{\delta}{3}-\frac{\delta^2}
{9\varepsilon}\,,\,\varepsilon+\frac{\delta}{3}+\frac{\delta^2}{9\varepsilon}\,,
\,1+\frac{\delta}{3}\right \}.\nonumber \end{eqnarray} Combined
with the results obtained from the diagonalization of
$\tilde{M}_{\rm eff}$, they lead to:
\begin{eqnarray}
\tan \theta_{12}=1-\frac{2}{3}\frac{\delta}{\varepsilon}+
\frac{2}{9}\frac{\delta^2}{\varepsilon^2}\;,\;\varepsilon =
\sqrt{\tan \theta_{12}\frac{\Delta m_{21}^2}{\Delta
m_{32}^2}}\;\,, \label{t} \nonumber
\end{eqnarray}
\begin{eqnarray}\delta =
\frac{3}{2}\left(1-\sqrt{2\tan\theta_{12}-1}\right) \sqrt{\tan
\theta_{12}\frac{\Delta m_{21}^2}{\Delta m_{32}^2}} \,,
\label{param3}
\end{eqnarray}
which replace Eqs.~(\ref{param4}), whereas Eqs.~(\ref{3}) remain
valid in this case.

Since $M_l$ is not exactly of the democratic form, the matrix $W$
in $U_l=FW$ deviates slightly from the unit matrix due to the
nonzero values of $m_e$ and $m_{\mu}$. Therefore, we obtain for
$|\delta|\aprle |\varepsilon| \ll 1$:
\begin{eqnarray}
\frac{\sin^2 2\theta_{12}}{\sin^2 2\theta}=1-\frac{4}{3}\frac{m_e}
{m_\mu}\frac{\cos 2\theta}{1-\cos
2\theta}\left(1+\frac{2}{3}\delta \right)\,, \label{S12a}
\end{eqnarray}
\begin{eqnarray}
\sin^2 2\theta_{23}=\frac{8}{9}\left(1+\frac{2}{3}\delta \right)
\left[1+\frac{m_\mu}{m_\tau}(1-3\delta )\right]\,, \label{S23a}
\end{eqnarray}
\begin{eqnarray}
\sin\theta_{13}=U_{e3}=-\frac{\varepsilon\delta}{3\sqrt{2}}-
\frac{\sqrt{2}}{3}\frac{m_e}{m_\mu}\left(1-\frac{\delta}{3}\right)\,.
\label{S13a}
\end{eqnarray}
In the limit $m_e, m_\mu \to 0$ the corresponding expressions of
Eqs.~(\ref{3}) and (\ref{t}) are recovered.

When $|\varepsilon|\ll |\delta| \ll 1$, one has: \be \sin^2
2\theta_{12}=4\left[\frac{3}{4}\frac{\varepsilon}{\delta}-\frac{1}{3}
\frac{m_e}{m_\mu}\left(1+\frac{2}{3}\delta\right)\right]^2 \,,
\label{S12b} \ee whereas $\sin^2 2\theta_{23}$ and $U_{e3}$ are
again given by Eqs.~(\ref{S23a}) and (\ref{S13a}). Notice that the
new contributions coming from nonzero $m_e$ and $m_{\mu}$ tend to
increase the values of $\sin^2 2\theta_{23}$, bringing it closer
to the Super-Kamiokande best fit value. They also increase
significantly the value of the mixing parameter $|U_{e3}|$.\\
\indent We have performed exact numerical diagonalizations of the
light neutrino effective mass matrix for each solution of the
solar neutrino problem. All the results are within the allowed
range of the experimental data for solar and atmospheric neutrinos
(see table 1)\cite{data}. Our analytical expressions also give
very accurate results when compared with the exact ones.\\ \indent
A simple renormalization group analysis of our {\em Ans\"{a}tze} shows
that these solutions are stable against quantum corrections at
one-loop level.\\

 \indent In conclusion, we have proposed a viable structure for
 the lepton mass matrices based on the extended democracy
 hipothesis, which, in the absence of right-handed neutrinos,
 would lead to small mixing in the leptonic sector, similar to what happens for quarks.
 The large mixing required by the atmospheric neutrino data appears naturally as a result of
 the seesaw mechanism combined with a strong hierarchy in the
 right-handed neutrino mass spectrum. By changing the values of
 the right-handed neutrino masses, we are able to reproduce all
 the possible solutions of the solar neutrino problem.

\end{document}